\documentclass[11pt]{article}
\usepackage{moriond,epsfig}

\def\Journal#1#2#3#4{{#1} {\bf #2}, #3 (#4)}

\def\NIMA{{\em Nucl. Instrum. Methods} A}

\def\be{\begin{equation}}
\def\ee{\end{equation}}
\def\bea{\begin{eqnarray}}
\def\eea{\end{eqnarray}}

\begin{document}
\vspace*{4cm}
\title{MASSES, LIFETIMES AND DECAYS OF B HADRONS AT THE TEVATRON}

\author{ L. VACAVANT }

\address{Lawrence Berkeley National Laboratory, Physics Division, \\
1 Cyclotron Road Bldg 50B-6222,\\
Berkeley, CA 94720-8167, USA\\
lvacavant@lbl.gov}

\maketitle\abstracts{
The latest results in B physics from the CDF and D0 experiments 
at the Tevatron are 
presented, including inclusive $b$ lifetime measurement, exclusive 
lifetime measurement of the $B_s$.
Promising samples collected by CDF with its Secondary Vertex Trigger 
are shown as well.
}

\section{Triggers and data samples}

Even though the $b\bar{b}$ cross-section at the Tevatron is five orders of magnitude higher than at the $\Upsilon_{4s}$ or at the $Z^0$ pole, 
adequate trigger strategies are required to extract $B$ physics since the 
total inelastic cross-section is about a thousand times larger than the 
$b\bar{b}$ one.

Both CDF~\cite{cdf} and D0~\cite{d0} at Run II are using a di-muon trigger to collect large samples of $J/\Psi$. The acceptance for the muons in this case varies between $|\eta|\leq 1$ (CDF) and $|\eta|\leq 2$ (D0), 
while the implicit thresholds on the muon $p_T$ are about 1.5, 2-2.5 and 3.5 GeV/c for respectively CDF, D0 forward and D0 central regions.
In addition, CDF is operating a new trigger system, 
the Secondary Vertex Trigger~\cite{svt}, to obtain samples with high heavy flavors contents.
The SVT combines the track parameters measured at the trigger first level in the Central Outer Tracking chamber with the axial hits from the Silicon VerteX detector, allowing to measure precisely the impact parameter of charged track at the level two of the trigger.
The SVT is used to collect two additional samples for B physics. 
A sample enriched in semi-leptonic decays is collected by requiring one SVT track ($|\eta|<1$, $p_T>2$ GeV/$c$, $d_0>120 \mu$m) in conjunction with a lepton ($p_T>4$ GeV$/c$, 
$|\eta^\mu|\leq 1$, $|\eta^{e}|\leq 1.5$).
Requiring two SVT tracks leads to a sample enriched in hadronic decays.

The following results are based on 60-70 pb$^{-1}$ of data collected between February and December 2002 for CDF, and on 40-45 pb$^{-1}$ between August 2002 and January 2003 for D0.

\section{Inclusive lifetime measurements}

\begin{figure}\centering
\epsfig{figure=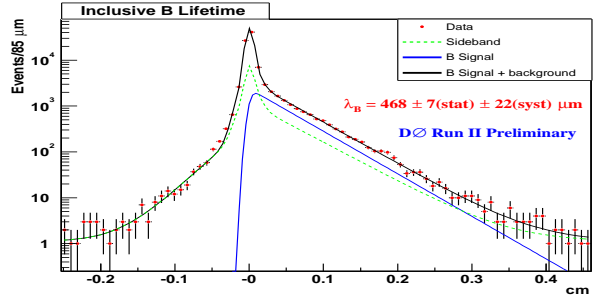,height=4cm,width=0.5\linewidth}
\caption{Lifetime distribution for inclusive $b$ decays.}
\label{fig:ctd0}
\end{figure}

The measurement of the inclusive $b$ hadron lifetime 
($B\rightarrow J/\Psi X$) is based 
on the transverse decay length $L=\beta\gamma ct = \frac{p_B}{m_B}ct$, 
which is the distance between the production point and the decay point,
measured in the plane transverse to the beam axis and projected onto the 
$B$ transverse momentum.
Since the $b$ hadron is not fully reconstructed, the correction factor 
$K=p_{J/\Psi}/p_B$ is taken from Monte-Carlo simulations.
The lifetime distribution measured by D0 is shown on figure~\ref{fig:ctd0}.
A large fraction of the sample consists of prompt $J/\Psi$, produced 
directly in the $p\bar{p}$ collision: 
those are used to study the resolution function which is the main source of systematic errors.
The remaining contributions to the systematic errors come from the $K$ factor, the modeling of the background and the alignment of the silicon.
The inclusive $b$ hadron lifetime is measured to be $1.561\pm 0.024 (stat.)\pm 0.074 (syst.)$ ps.
CDF performed a similar measurement in July 2002 with 18 $pb^{-1}$ of data, leading to $\tau = 1.526 \pm 0.034 (stat.) \pm 0.035 (syst.)$ ps.

\section{Exclusive lifetime measurements}

\begin{figure}
\begin{minipage}{0.49\textwidth}
\epsfig{figure=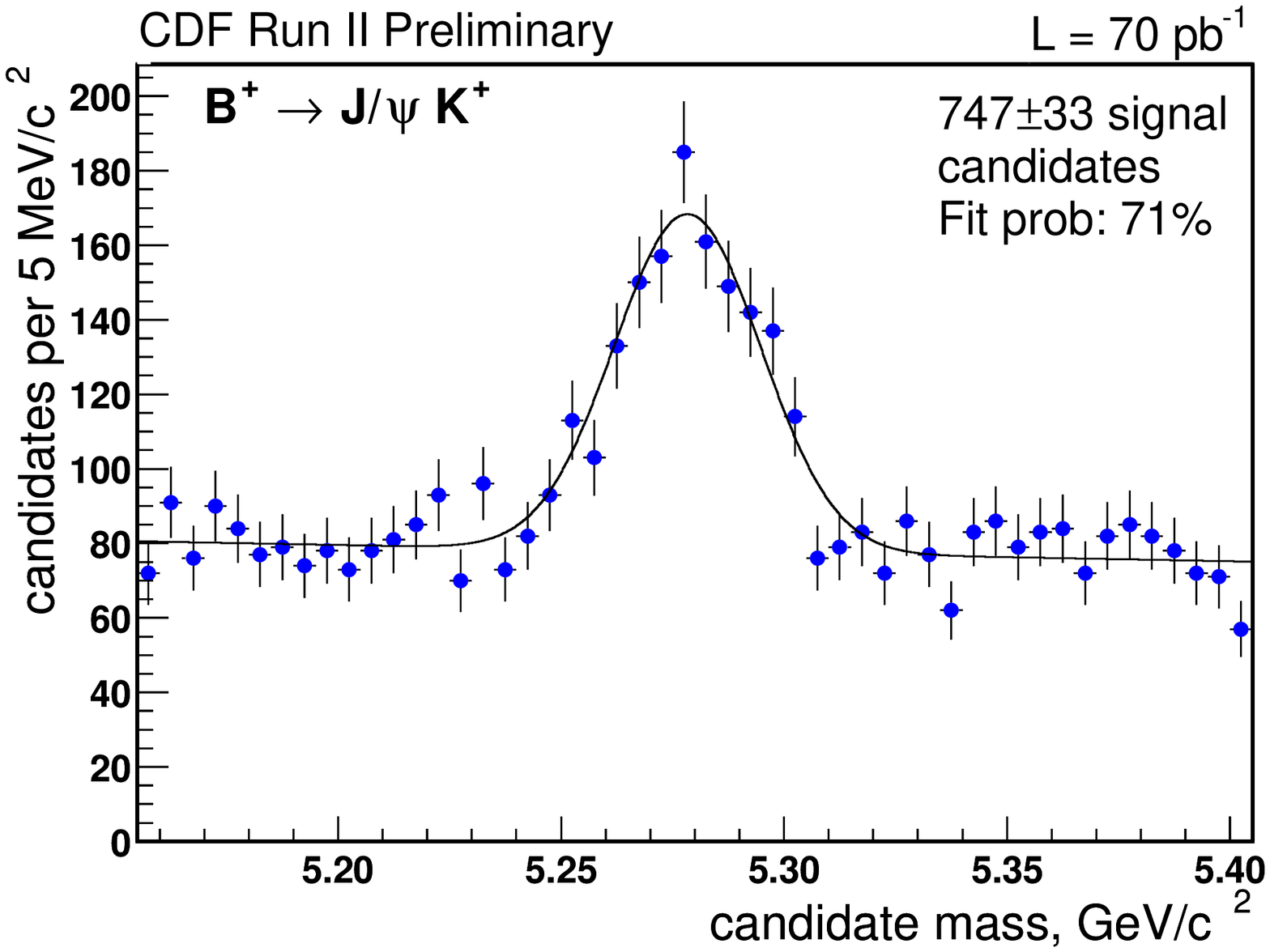,height=4cm,width=\linewidth}
\end{minipage}
\hfill
\begin{minipage}{0.49\textwidth}
\epsfig{figure=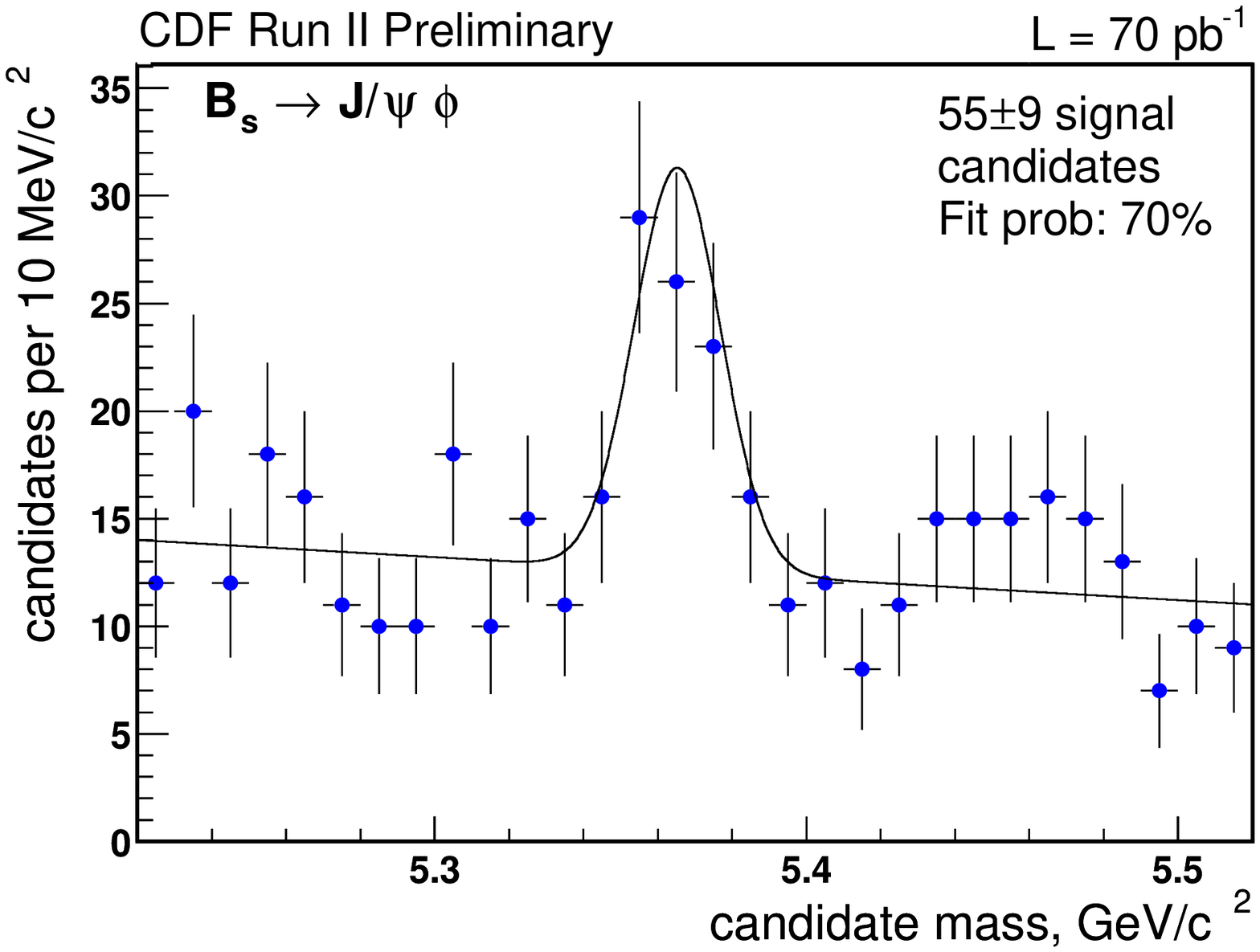,height=4cm,width=\linewidth}
\end{minipage}
\caption{Invariant mass spectrum of $B^+\rightarrow J/\Psi K^+$ (left) 
and $B_s\rightarrow J/\Psi \Phi$ (right) candidates.}
\label{fig:bex}
\end{figure}

\begin{figure}
\begin{minipage}{0.49\textwidth}
\epsfig{figure=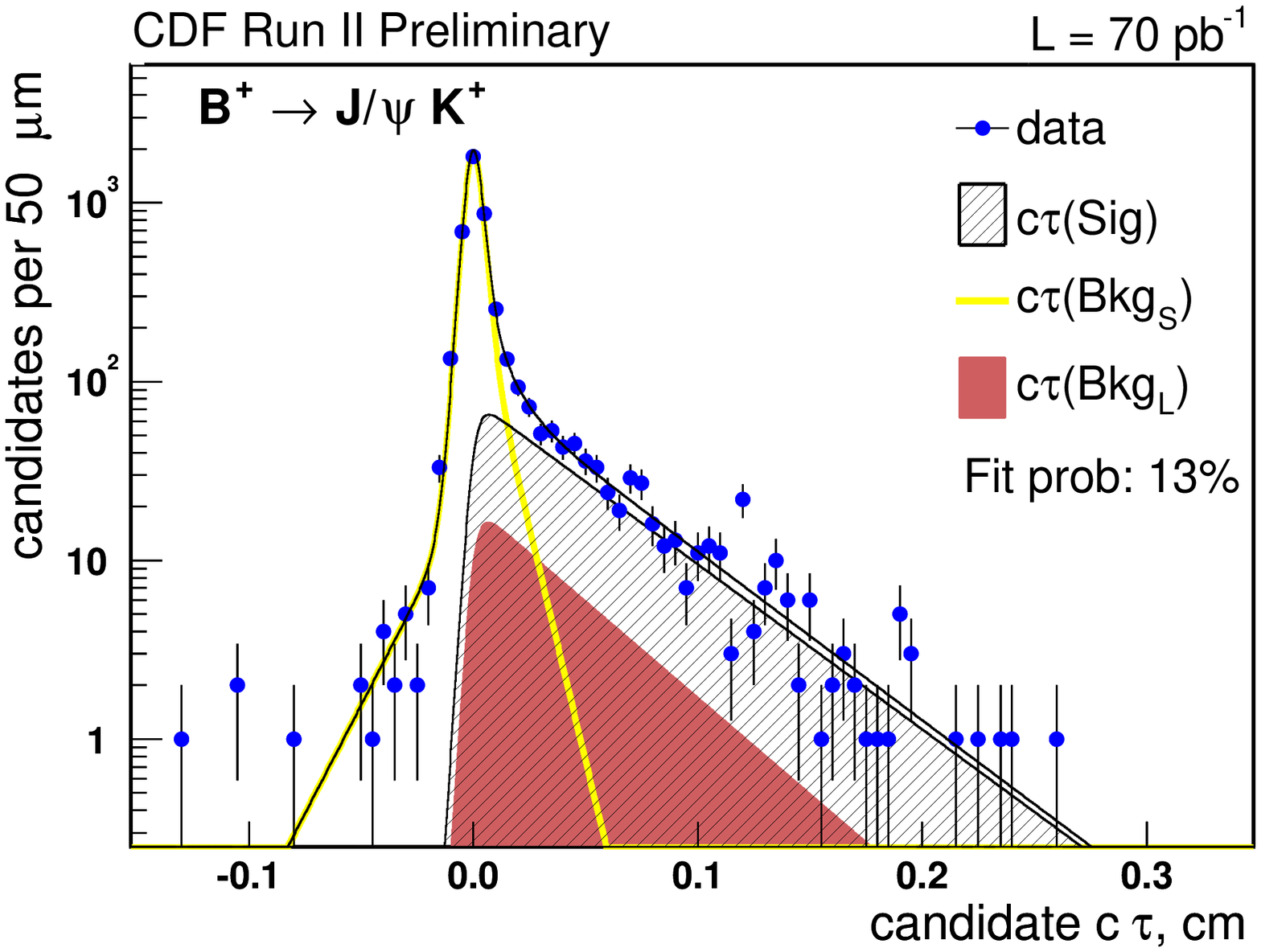,height=4cm,width=\linewidth}
\end{minipage}
\hfill
\begin{minipage}{0.49\textwidth}
\epsfig{figure=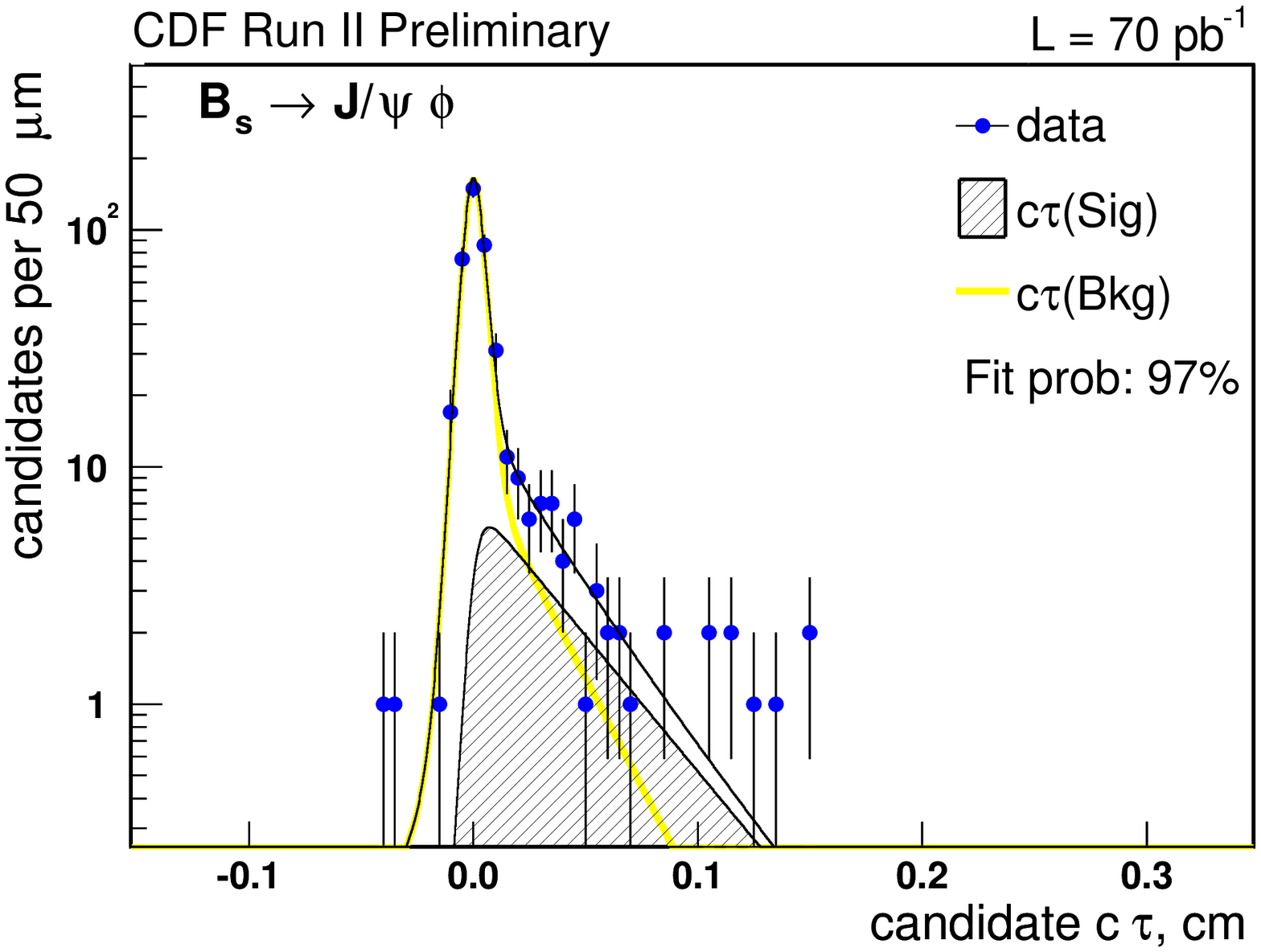,height=4cm,width=\linewidth}
\end{minipage}
\caption{Proper decay length distributions and fit results for 
the $B^+\rightarrow J/\Psi K^+$ (left) 
and $B_s\rightarrow J/\Psi \Phi$ (right) candidates.}
\label{fig:bexfit}
\end{figure}

CDF has extracted the lifetime of the $B^+$, $B^0$ and $B_s$ mesons from the 
following channels: $B^+\rightarrow J/\Psi K^+$, $B^0\rightarrow J/\Psi K^*$ 
and $B_s\rightarrow J/\Psi \Phi$.
The $B^+$ and $B^0$ analyses are used as a control sample to validate the techniques used for the $B_s$ lifetime measurement and to evaluate the systematic uncertainties.
Data are collected using the di-muon trigger for $J/\Psi \rightarrow \mu^+\mu^-$.
The invariant mass spectrum of the selected $B^+$ and $B_s$ candidates are shown on figure~\ref{fig:bex}.
An unbinned maximum likelihood fit is used to extract the lifetime where the mass and proper decay length are fitted simultaneously. 
The proper decay length distributions are shown on figure~\ref{fig:bexfit}, with the result of the fit overlaid.
The results for the lifetimes are:
$$\begin{array}{ll|}
 \tau_{B^+} & =  1.57  \pm 0.07 \; (stat.)  \pm 0.02 \; (syst.) \mbox{ ps} \\
 \tau_{B^0} & =  1.42  \pm 0.09 \; (stat.)  \pm 0.02 \; (syst.) \mbox{ ps} \\
 \tau_{B_s} & =  1.26  \pm 0.20 \; (stat.)  \pm 0.02 \; (syst.) \mbox{ ps} \\
\end{array}\hfill
\begin{array}{ll}
 \tau_{B+} / \tau_{B^0} & =  1.11  \pm 0.09\\
 \tau_{B_s} / \tau_{B_d} & =  0.89  \pm 0.15\\
\end{array}$$

\section{Signals for $\Lambda_b$}

\begin{figure}
\begin{minipage}{0.49\textwidth}
\epsfig{figure=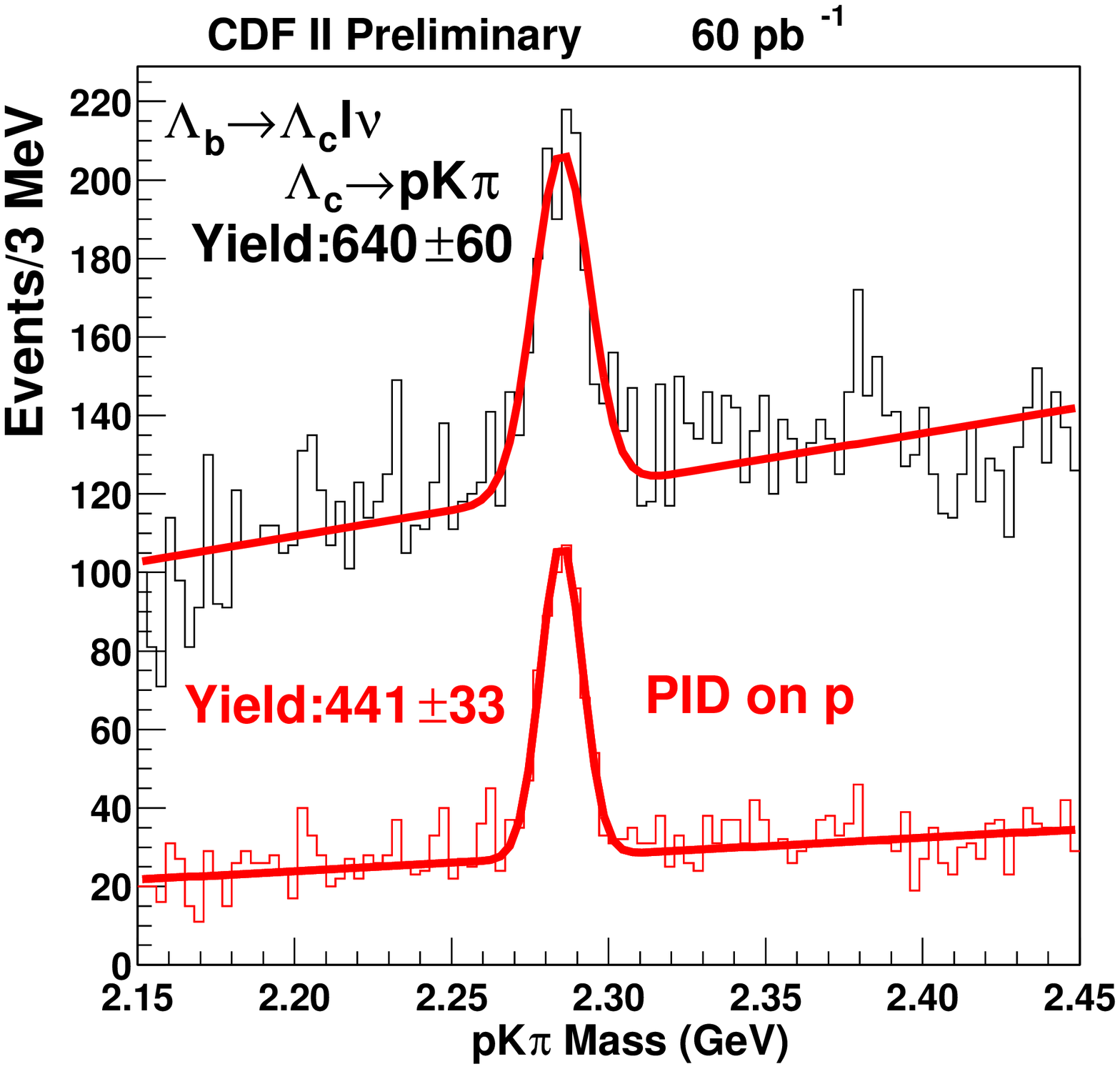,height=5cm,width=\linewidth}
\end{minipage}
\hfill
\begin{minipage}{0.49\textwidth}
\epsfig{figure=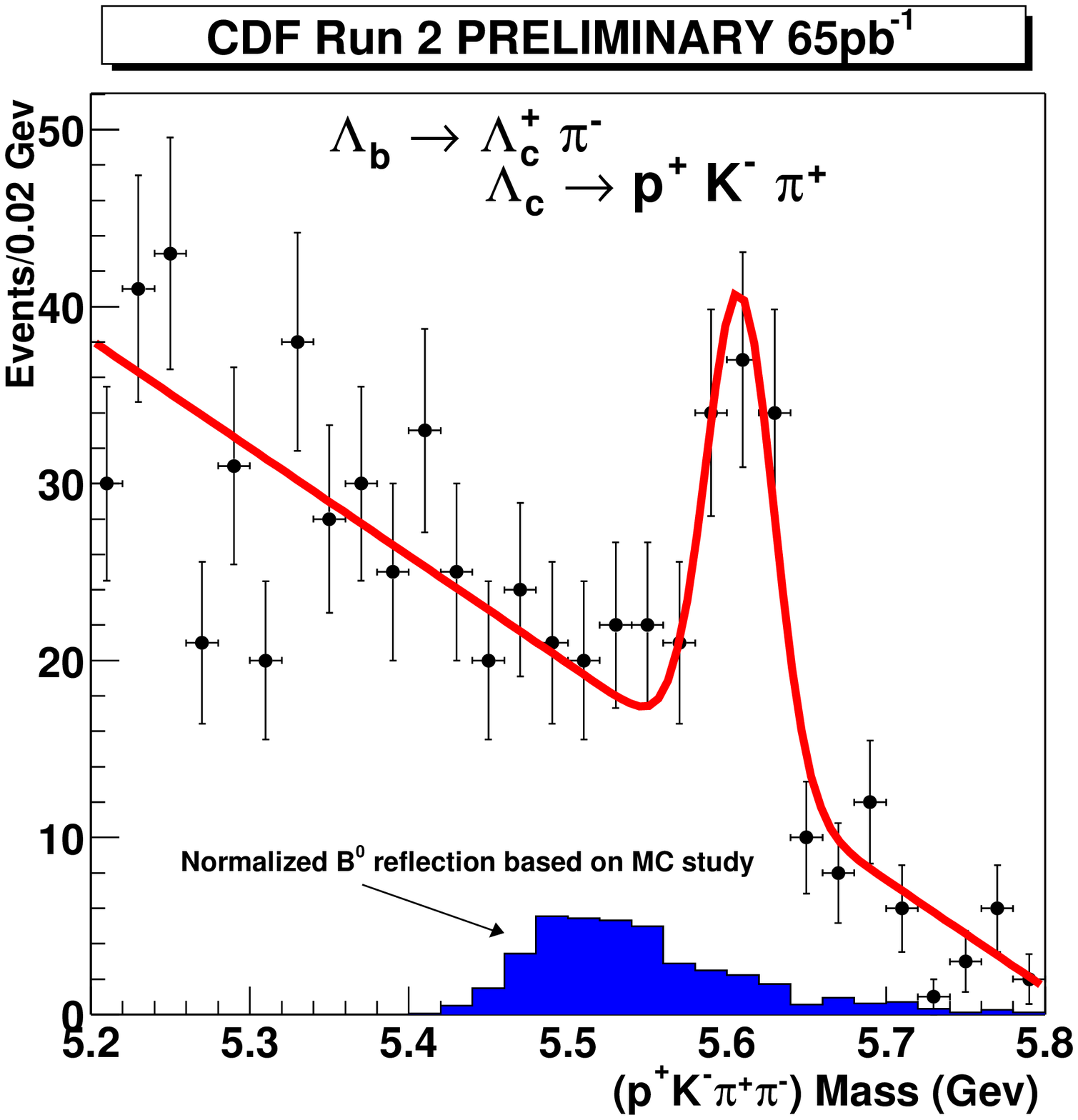,height=5cm,width=\linewidth}
\end{minipage}
\caption{$\Lambda_b$ signals: invariant mass distribution of 
$\Lambda_c$ candidates in the lepton+SVT sample (left) and of 
the $\Lambda_b$ fully reconstructed candidates (right).}
\label{fig:lb}
\end{figure}

The lifetime and branching ratios of the $\Lambda_b$ are poorly known and the experiments at the Tevatron have the unique opportunity to improve our knowledge in this field.
CDF has collected samples of $\Lambda_b$ already significant in size using the three trigger paths described above. 
Measurements based on those samples are on-going.
Figure~\ref{fig:lb} shows the events collected using the lepton+SVT trigger and the events collected with the two-track trigger. 
In the former case, the substantial gain in signal over noise by applying particle identification cuts on the proton is shown: 
in this momentum range, the proton identification relies mostly on the new Time-Of-Flight sub-detector. 
The latter sample is the largest sample of fully reconstructed 
$\Lambda_b$ hadronic decays.

\section{Signals for $B_s$ with the SVT}

\begin{figure}
\begin{minipage}{0.32\textwidth}
\epsfig{figure=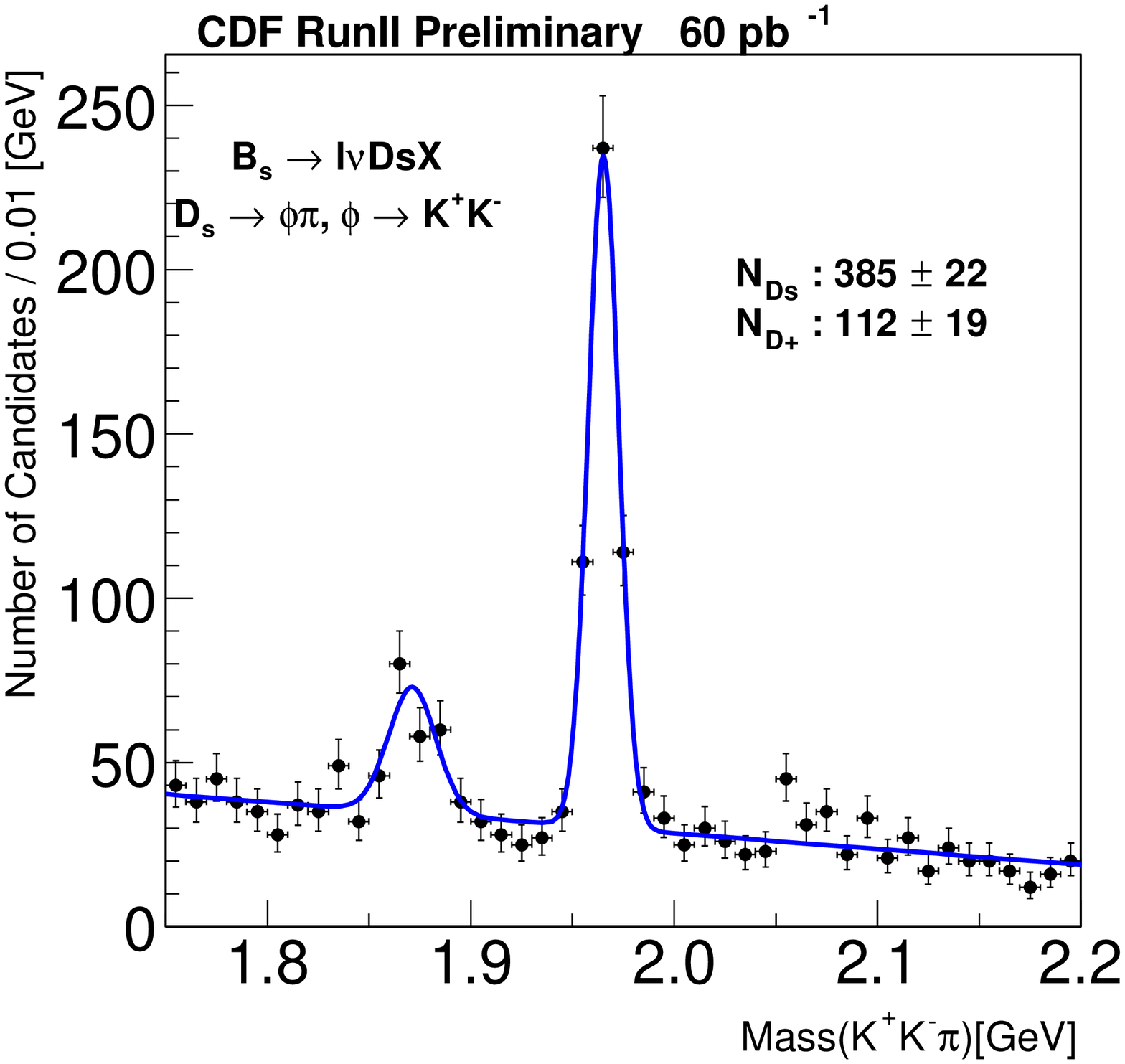,height=5cm,width=\linewidth}
\end{minipage}
\hfill
\begin{minipage}{0.32\textwidth}
\epsfig{figure=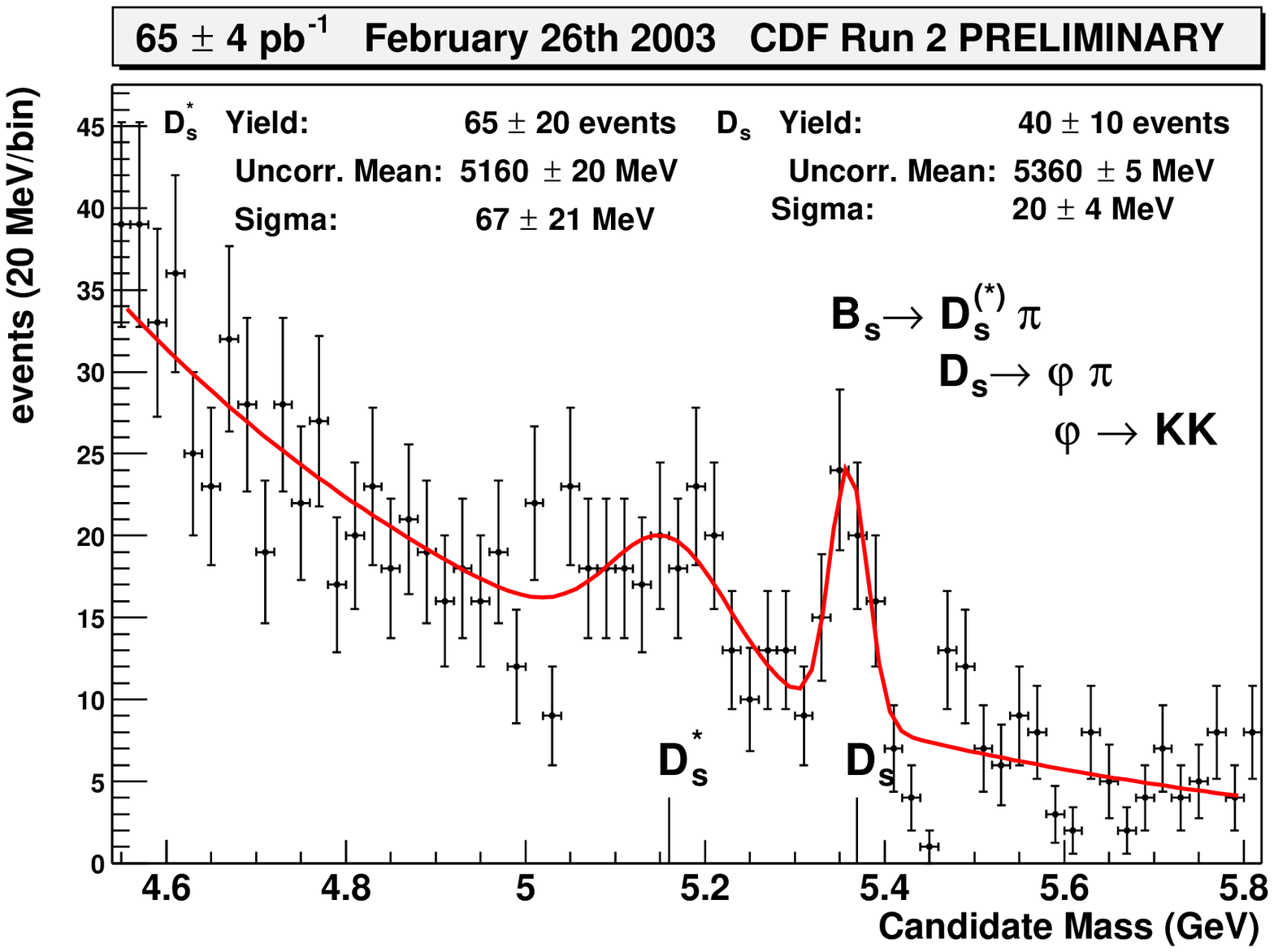,height=5cm,width=\linewidth}
\end{minipage}
\hfill
\begin{minipage}{0.32\textwidth}
\epsfig{figure=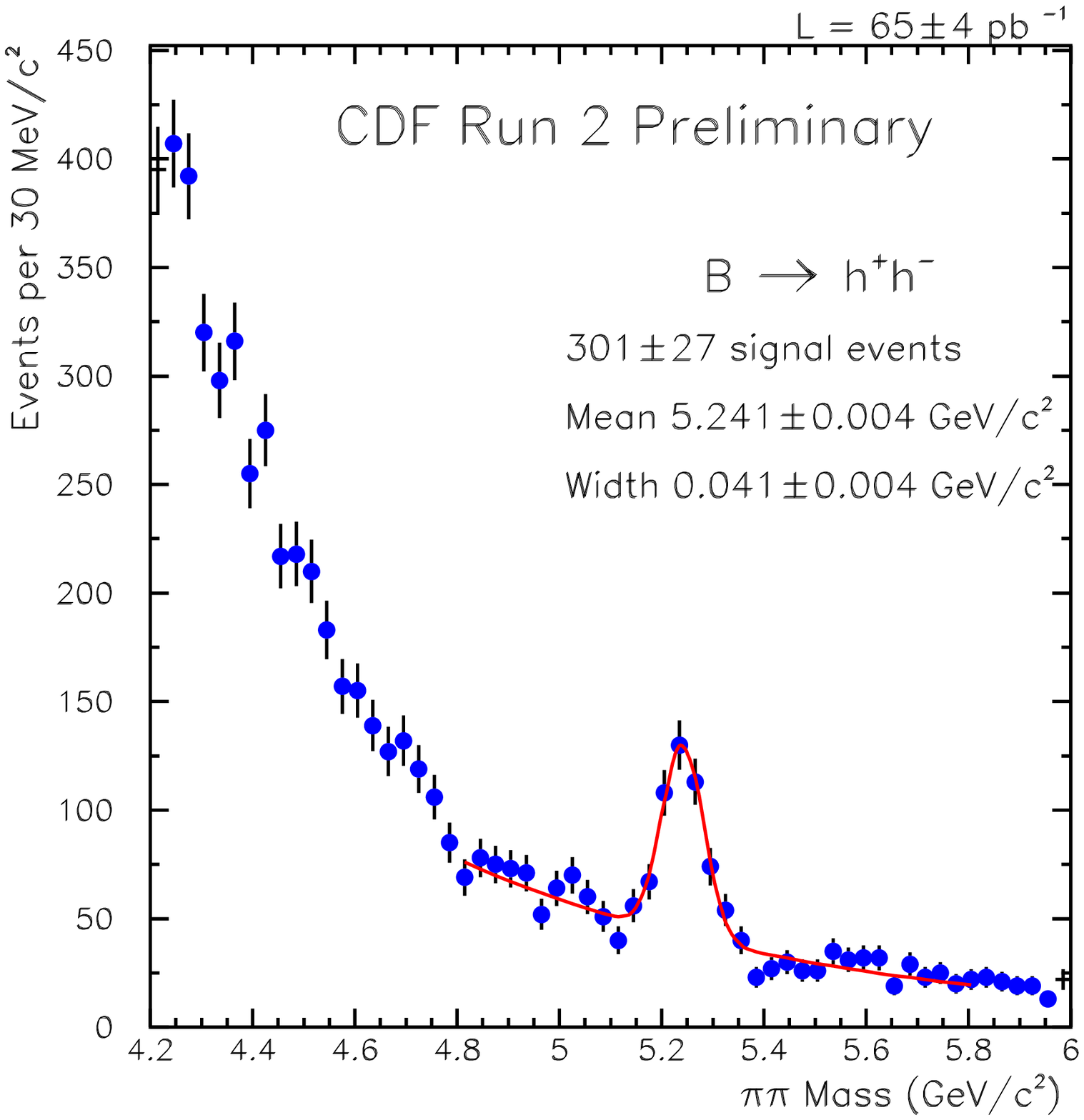,height=5cm, width=\linewidth}
\end{minipage}
\caption{Signals obtained with SVT: $B_s$ semi-leptonic decays (left),
$B_s$ fully hadronic decays (middle) and
charmless hadronic decays (right).}
\label{fig:bs}
\end{figure}

The observation and measurement of $B_s$ oscillations is one of the major goals of the $B$ physics program at the Tevatron.
The left plot from figure~\ref{fig:bs} shows the semi-inclusive 
$B_s$ sample 
($B_s\rightarrow D_s l \nu X \rightarrow [ \phi \pi ] l \nu X \rightarrow [  [ K^+ K^- ] \pi ] l \nu X $) 
 collected with the lepton+SVT trigger by CDF, with $385\pm22$ events.
 This sample is being used for lifetime measurement $(\tau(B_s)/\tau(B_d))$ and could be used for $B_s$ mixing for moderate values of $x_s$.
 However the golden sample for this measurement is the one based on the fully reconstructed channels, allowing to resolve fast oscillations:
 $  B_s  \rightarrow  D_s^{(*)-} \pi^+ 
 ,\;\;  B_s  \rightarrow  D_s^{(*)-} 3\pi  
 ,\;\;  B_s  \rightarrow  D_s^{(*)-} D_s^{(*)+} $
 Thanks to the SVT, CDF has observed $40\pm10$ $D_s\pi$ and 
 $65\pm20$ $D_s^* \pi$, shown on the middle plot of figure~\ref{fig:bs}.
 More channels are being added and the trigger configuration is being improved to maximize those yields.

\section{Charmless hadronic decays}

The two-track trigger gives CDF access to charmless hadronic decays, 
a first at a hadron collider.
A total of $301\pm27$ of such events has been collected, their invariant mass distribution is shown on figure~\ref{fig:bs}, right plot.
Studies are going on to disentangle the four contributions: 
$B^0\rightarrow K^+\pi^+$, $B^0\rightarrow \pi^+\pi^-$, 
$\bar{B}_s^0\rightarrow K^+K^-$,
$\bar{B}_s^0\rightarrow K^+\pi^-$, using the kinematics and the 
particle identification power provided by the measurement of the energy loss by ionisation in the COT.

\section{Conclusion}

The CDF and D0 experiments at the Tevatron are back online studying 
$B$ physics. 
Unique measurements are being addressed while collecting more data.
The ability to trigger on displaced vertices has already opened new 
opportunities and is very promising for future measurements.

\section*{Acknowledgments}

We would like to acknowledge the work of all the CDF and D0 collaborators, as well as the work of the people from the Tevatron.
Thanks to the Moriond organizers for this pleasant and profitable conference.

\end{document}